\documentclass[10pt,a4paper]{article}
\usepackage[utf8]{inputenc}
\usepackage[english]{babel}
\usepackage{amsmath}
\usepackage[pdftex]{graphicx}
\usepackage{epstopdf}
\usepackage{caption}
\usepackage{multicol}
\usepackage{subcaption}
\usepackage{times}
\usepackage{hyperref}
\usepackage{float}
\usepackage{indentfirst}
\usepackage{geometry}
\usepackage{array}        

\begin{document}

\begin{center}
\fontsize{14}{16}\selectfont
\noindent \textbf{CHARACTERISTICS OF ELECTROMAGNETIC RADIATION AND THE INTERFERENCE PROTECTION ON A SMALL-SIZED DIRECT-ACTING ELECTRON ACCELERATOR WITH A PLASMA OPENING SWITCH}
\end{center}

\begin{center}
\fontsize{12}{14}\selectfont
\noindent \textbf{D.V. Vinnikov, O.M. Ozerov, V.V. Katrechko, V.I. Tkachov, S.V. Marchenko, V.V. Ehorenkov,}

\noindent \textbf{S.A. Petrenko, B.O. Brovkin, O.V. Manuilenko, I.N. Onishchenko}
\end{center}
\begin{center}
\fontsize{10}{12}\selectfont
\noindent \textbf{\emph{National Science Center "Kharkiv Institute of Physics and Technology", Kharkiv, Ukraine}}

\textbf{\emph{E-mail: onish@kipt.kharkov.ua}}
\end{center}

An experimental study of the parameters of electromagnetic and X-ray radiation was carried out on a small-sized direct-acting electron accelerator with inductive storage and a plasma opening switch. Frequency spectra have been determined, and the diagnostics of the power of the microwave component of the spectrum has been tested. An analytical comparison of radiation signals and power signals has been carried out. A method for determining the directivity pattern using a broadband power sensor has been proposed. An anechoic chamber and a reflective screen were used to suppress interference.

\begin{flushleft}
\fontsize{10}{12}\selectfont
 \end{flushleft}

\columnseprule=0pt\columnsep=24pt
\begin{multicols}{2}

\begin{center}
\fontsize{11}{12}\selectfont
\noindent \textbf{1. INTRODUCTION}
\end{center}
\par A small-sized direct-acting electron accelerator (DAEA) with inductive energy storage (IES) and a plasma opening switch (POS) is a high-voltage and high-current unit that serves as a source of X-ray (XR) and electromagnetic radiation (EMR).

The simplicity of maintenance and design of the accelerator, its small size, the absence of the need to use bulky Arkadyev-Marx generators as a high-voltage source, the exclusion of oil- or water-filled forming lines and magnetic systems for focusing the electron beam make these types of accelerators competitive and attractive for use as mobile operational devices in various industries, including critical infrastructure [1, 2].

The DIN-2K unit consists of an electric circuit of plasma guns (PG) and the main discharge circuit that provides energy storage [3-8]. Both circuits are pulse current generators, where the voltage switching in the range of 6 to 10 kV is carried out by air-controlled dischargers. These discharges are sources of electromagnetic radiation in a wide frequency spectrum that can distort useful signals. The propagation of interferences occurs both in free space and along the grounding circuits, and as a matter-of-fact appropriate measures should be taken to ensure the protection of the receivers of useful signals and analog-to-digital converters (ADC).

An important scientific task that is performed using DAEA DIN-2K accelerator consists in improving already known interference protection measures and testing new measures, capable of suppressing or eliminating the effect of interferences generated by the sources of pulsed current and voltage and without these measures the operation of the unit is impossible. At the same time, it is necessary to ensure correct diagnostics of useful signals whose source is an electron beam and a virtual cathode [9-12]. Thus, the issue of ensuring measures to provide interference protection is an urgent problem [13], which in our case was solved by performing the following tasks:

\noindent 1. Identifying interference sources.

\noindent 2. Recording the interference by diagnostic tools.

\noindent 3. Implementing engineering and technical solutions to eliminate interference.

\noindent 4. Verifying the quality of the measures taken to reduce and eliminate electromagnetic interference (EMI).

\noindent 5. Recording useful signals in the newly created conditions that improve the interference protection.

The purpose of this research is to study the sources of electromagnetic radiation and improve already available and newly developed methods and tools that provide protection of tested useful signals generated by DAEA and exposed to electromagnetic interference, creating thus conditions for the unhindered operation of diagnostic tools, and ensuring the recording of useful pulses with reduced interference.

\begin{center}
\fontsize{11}{14}\selectfont
\noindent \textbf{2. DESCRIPTION OF THE UNIT OPERATION AND DIAGNOSTICS}
\end{center}
\par The operation of accelerator consists in the sequential activation of its electrical circuits at the present delay time. Fig. 1. gives a general view of the DAEA located in the anechoic chamber.
Using the remote-control desk, the operator selects the discharge voltage for the electrical circuits of the PGs in the range of 6 kV to 18kV that are triggered via the P2 discharger, and for the main electrical circuit from 15 kV to 50 kV, which is activated by D1 discharger. The delay time is preset using the time delay unit, varying in the range of 0 to 50 $\mu$s.

After the circuits are triggered, the energy stored in a low-inductance pulse capacitor of a 50 kV/3 $\mu$F type is directed into the coaxial accelerating electrode systems 3, 4, where, taking into account the inductance of the electrodes and the time-varying current flowing through them, the energy storage (IES) 1 occurs, and the release of it is realized during the opening of the current loop (POS) 6 that  is created in the plasma injected into the volume by the guns, closing the cathode and anode, see Fig. 2. After the POS is opened, a sharp drop in the conductivity of the gap filled with plasma is observed, approximately by two orders of magnitude, resulting in a jump-like increase in the discharge voltage of the main circuit, within a time  

\begin{figure}[H]
  \centering
  \includegraphics[width=0.45\textwidth]{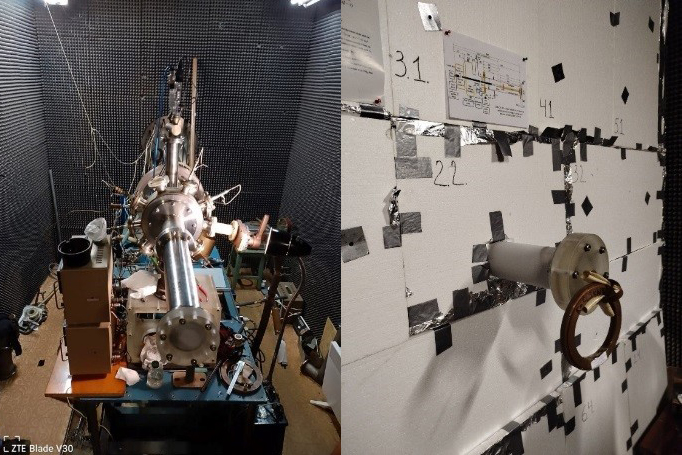}\\
  \caption{\emph{General view of the DAEA DIN-2K with no interference-absorbing screen (photo on the left) and with a screen.}}\label{fig:image_label}
\end{figure}

\noindent of 20 to 100 ns. When the critical voltage is exceeded, attaining more than 120 kV, these processes result in the formation of explosive emission of electrons from the end of the tubular cathode and, accordingly, in the formation of an electron beam 7. Provided that the beam current exceeds the critical vacuum current at a distance approximately equal to the size of the gap between the cathode and the anode grid 5, a negative volume charge, or the so-called virtual cathode 8 (VC) can be formed behind the grid. The space-time oscillations of the VC relatively the vacuum diode 2 (VD) excite electromagnetic oscillations in the chamber that are emitted into the open space. These useful oscillations must be separated from the electromagnetic radiation generated in the unit during its operation.

Induced interferences affect diagnostic tools that can be located both inside and outside the chamber. The beam current that is the power source for the virtual cathode is measured using a disk collector with a diameter of 70 mm. The beam current and the current of the main discharge circuit are recorded by Rogowski coils with a bandwidth of 1.6 MHz to 994.7 MHz. The charging voltage is determined by kilovoltmeters of a C-196 type, a North Star High voltage divider of a PVM-12 type, 1:1000, 80 MHz, and a North Star High voltage divider of a PVM-1 type, 1:1000, 120 MHz, 60 kV. The opening voltage is measured by a capacitive voltage divider calibrated using a certified PVM-1 voltage divider, and by a non-invasive calculation method based on the data obtained from oscillograms during the POS opening [14, 15]. Rogowski coils intended for measuring currents, an X-ray sensor, and receiving antennas that record the spectral characteristics of the radiation and its power are located outside the chamber.
The measured frequency spectrum was recorded using an Agilent infiniium 54855A DSO, 6 GHz, 10 GSa/s oscilloscope, connected to a 2.5cm long rod antenna via the first receiving channel and via the second channel to a waveguide combined with a coaxial-waveguide transition with a cutoff frequency of 2.1 GHz and an available microwave diode 8474B Detector, Coaxial, 0.01 – 18 GHz.

Microwave power was recorded in a wireless mode by a broadband power sensor U2021XA 50 MHz to 18GHz USB Peak Power Sensor additionally equipped with a horn receiving antenna and a coaxial-wave adapter. Trapezoidal horn antenna had a cutoff frequency of 2.1GHz. The frequency of microwave radiation was controlled by the size of the anode-cathode gap of the vacuum diode and the discharge voltage of PCG-1. The amount of X-ray radiation was determined using the dosimeters of a DK-02, ID-02 type and gallium arsenide-based scintillation crystals. Oscilloscopes used were Siglent SDS 2204X, 200 MHz and OWON XDS3104E, 100 MHz.

Devices that could be affected by interference include a pulse triggering unit of 18 kV, a delay unit based on a G5-15 pulse generator with a delay time of 100 ns-500 $\mu$s and with a time step of 100 ns.
The vacuum control was provided by VIT-2 devices. During the experiments, the working vacuum was not worse than 2·10-5 Torr.

The primary absorption of interference propagating around the volume of the measuring chamber occurs on its walls. It is an anechoic chamber with ferrite absorbers of pyramidal shape 12, 60 mm long and it simulates free space, which provides attenuation of interference from 100 MHz to 45 GHz with a reflection coefficient of -10 dB [17]. Figure 1 also shows the sources of unwanted interference and the sources of useful signals 10. Since there are always such directional interferences that fall on the receivers of the measuring equipment located in front of the wall of the anechoic chamber, a special gapless, prefabricated, and lightweight metal screen 11 was arranged to separate and absorb this part of the interference from the dischargers. EMI generated during explosive emission and opening of the POS is shielded by a grounded anode grid. The grid step is 5x5 mm, which allows us not to pass all the frequencies up to 6 GHz. Geometric transparency is 70-75\%.

Diagnostic equipment, namely oscilloscopes, are separated from the general network and are powered by uninterruptible power supply sources Smart Logic power – 6000 PRO, which prevents interference from entering through the common grounding system.

\begin{center}
\fontsize{11}{14}\selectfont
\noindent \textbf{3. ACCELARATOR PARAMETERS}
\end{center}

The main parameters of the DAEA that affect the spectrum and intensity of interference and useful radiation are the discharge voltage, the induced voltage and the beam current. The effect of the electrical parameters of the circuits on the output characteristics of the accelerator, namely the current amplitude during energy storage in the IES, the value of the induced voltage, the electron beam current, the level of X-ray radiation are considered in papers [14-16]. Numerical modeling of the dynamics of the POS and the conditions for its formation is considered in papers [18-23]. The intensity and width of the spectrum of the useful signal are controlled by the discharge voltage of the main circuit, by the voltage of the circuit of plasma guns and the size of the vacuum diode gap between the cathode and the anode grid.
\end{multicols}

\begin{figure}[H]
  \centering
  \includegraphics[width=0.9\textwidth]{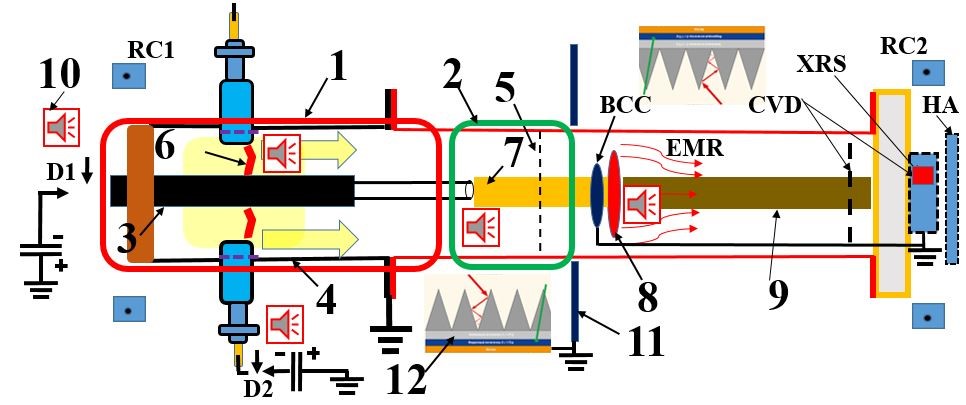}\\
  \caption{\emph{1 – IES; 2 – vacuum diode; 3 – cathode; 4 – anode chamber housing; 5 – anode grid; 6 – POS, 7 – electron beam, 8 – virtual cathode; 9 – electron beam after passing through the virtual cathode; 10 – marking of the main sources of interference and useful signals; 11 – metal solid screen; 12 – ferrite radio-absorbing plates of the anechoic chamber; BCC – beam current collector; EMR – targeted electromagnetic radiation; CVD– capacitive voltage divider; XRS– X-ray sensor; RC1, RC2 – Rogowski coils; HA – receiving horn antenna; D1, D2 – controlled air dischargers.}}\label{fig:image_label}
\end{figure}

\columnseprule=0pt\columnsep=24pt
\begin{multicols}{2}
\par The beam current varies in the range of 5 to 10kA, increasing with an increase in the discharge voltage of the main circuit and decreasing with an increase in the distance of the current storage from the anode grid of the vacuum diode.

The parameter that additionally confirms the presence of an electron beam is the formed X-ray radiation. The dependence of the X-ray dose and the maximum beam current (which was measured at 30 mm from the anode grid) on the discharge voltage of the main circuit is shown in Fig. 3. It is seen that the X-ray dose is continuously increasing with an increase in the discharge voltage, which is associated with an increase in the number of high-energy electrons. At the same time, an increase in the beam current strength is stabilized at voltages exceeding 30 kV, which is associated with the limitation of electron emission due to the limited cathode surface area.

\begin{figure}[H]
  \centering
  \includegraphics[width=0.47\textwidth]{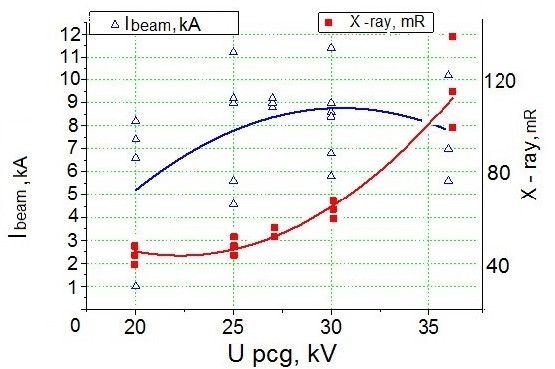}\\
  \caption{\emph{Dependence of the beam current strength and X-ray dose on the discharge voltage of the main circuit of the DAEA.}}\label{fig:image_label}
\end{figure}

Figure 4 shows a typical oscillogram of the full current with the current during the POS opening and the subsequent restoration of the oscillatory circuit.

\begin{figure}[H]
  \centering
  \includegraphics[width=0.5\textwidth]{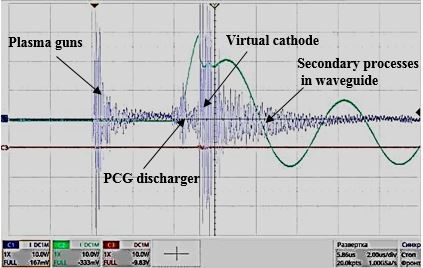}\\
  \caption{\emph{Current oscillogram, curve 1, and the interference signal, curve 2.}}\label{fig:image_label}
\end{figure}

Current oscillogram 1 demonstrates the moment of the beginning of an increase in current after the operation of the air-controlled D2 discharger of the main circuit, shown on curve 2. Then the moment of completion of the energy storage in the IES and the beginning of its release, which coincides with a rapid drop in current, about 400 kA/$\mu$s. After that, the current continues to propagate through the plasma formed by the guns and the electron beam in the vacuum diode, provided that not all the energy is released from the capacitor bank. Curve 2 is formed by a signal recorded by an air transformer in the circuit of plasma guns that is also capable of perceiving interferences formed by currents in other nodes of the unit, which helps us to form a holistic chronogram of physical processes.

Depending on the initial conditions, the main initial parameters of the accelerator are as follows: pulse duration of 50 to 200 ns, the beam energy varying in the range of 100 to 400 keV, the beam current of 5 to 11 kA, the beam diameter of 30 mm, the X-ray dose of 40 to 

\noindent 200 mR/pulse. The maximum value of the vector modulus of electric field at 0.5 m is 52.49 kV/m, at a discharge voltage of 25 kV and 8 kV on plasma guns. The maximum value of the vector modulus of magnetic field recorded in the air space at 0.5 m is 9.2 A/m, at the same voltage values.

\begin{center}
\fontsize{11}{14}\selectfont
\noindent \textbf{4. EXPERIMENTAL PART}
\end{center}

To determine the efficiency of the designed interference-proof screen, the intensities of the spectra were measured first without the screen and then with it. As can be seen in Fig. 5, a signal with a duration of about 1.5 $\mu$s arrives in the measuring antenna, consisting of a waveguide and a microwave diode, curve 1, after the plasma gun discharger is triggered, which corresponds to the duration of the discharge current front of plasma guns. The spectrum consists of frequencies up to 30 MHz. Further, after 3.5 $\mu$s, which corresponds to the delay time set on the delay unit, the controlled air discharger of main circuit is triggered, see. curve 2. After that, with a delay of 1.3 $\mu$s, which corresponds to the time of energy storage in the IES, and coincides with the rise of the front on the current oscillogram, see Fig. 4, curve 1, the POS is opened and a VC is formed, which generates a microwave spectrum, while the microwave diode is triggered and it is locked for a time of up to 150 ns, see Fig. 5, curve 1. The arranged interference protection screen (see Fig. 6) provided a significant reduction by -7 dB in the interference signal from the dischargers of the plasma guns and the main discharge circuit. It indicates a 5-fold reduction in the power output.

The rod antenna records an intense microwave signal with peak frequencies in intensity in the range of 3.5-5.0 GHz, see Fig. 7. Subsequently, secondary discharge processes occur, characterized by the restoration of conductivity and the restoration of the main oscillatory circuit.

\begin{figure}[H]
  \centering
  \includegraphics[width=0.47\textwidth]{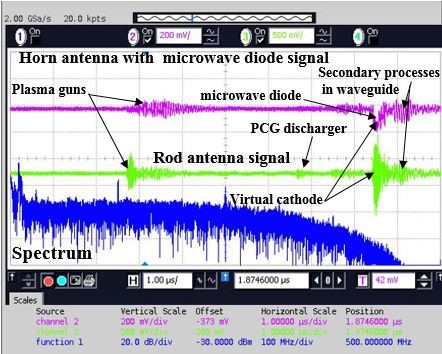}\\
  \caption{\emph{Chronogram of the signals prior to the arrangement of the protection screen.}}\label{fig:image_label}
\end{figure}

\begin{figure}[H]
  \centering
  \includegraphics[width=0.45\textwidth]{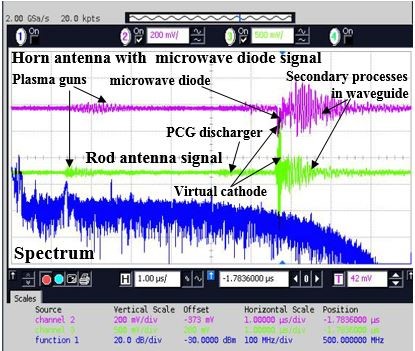}\\
  \caption{\emph{Signal chronogram after the arrangement of the protection screen.}}\label{fig:image_label}
\end{figure}

\begin{figure}[H]
  \centering
  \includegraphics[width=0.5\textwidth]{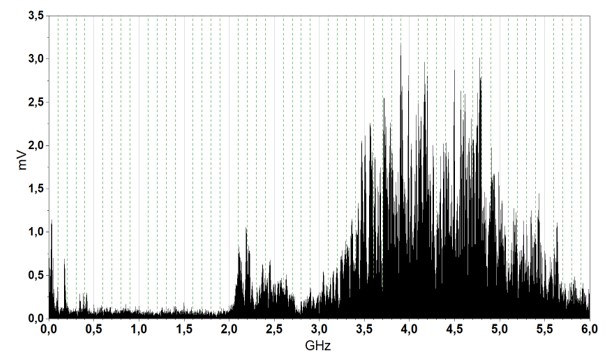}\\
  \caption{\emph{Fast Fourier Transformation of a Microwave Signal from VC.}}\label{fig:image_label}
\end{figure}

\begin{figure}[H]
  \centering
  \includegraphics[width=0.5\textwidth]{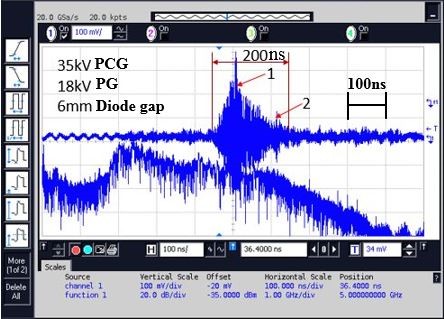}\\
  \caption{\emph{Signal with spectrum on a Horn antenna without a microwave diode.}}\label{fig:image_label}
\end{figure}

The useful microwave signals that were output from the waveguide and formed by the virtual cathode were recorded at the primary level.
When measuring the spectrum, the characteristic frequencies were recorded on the waveguide without a microwave diode, see Fig. 8.

\begin{figure}[H]
  \centering
  \includegraphics[width=0.5\textwidth]{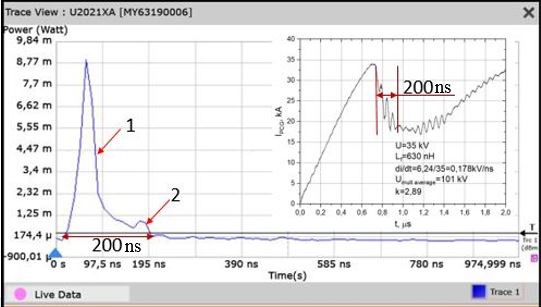}\\
  \caption{\emph{The shape of the signal of microwave power with the signal current oscillogram.}}\label{fig:image_label}
\end{figure}

By comparing the signals of electromagnetic radiation recorded by the antenna and the power of those signals in time, see Fig. 9, we can note the chronological coincidence of the duration of the microwave radiation with the duration of the power signal, as well as the coincidence of radiation bursts with power bursts, and the duration of the opening of the POS on the current oscillogram.

Using a power meter, we developed a method for evaluating the directivity pattern of a circular waveguide. A diagnostic stand was arranged with a platform on which a power sensor was installed every $11^\circ$, and up to 10 statistical measurements were taken at each angle, giving us an opportunity to determine the average power, see Fig. 10.

\begin{figure}[H]
  \centering
  \includegraphics[width=0.5\textwidth]{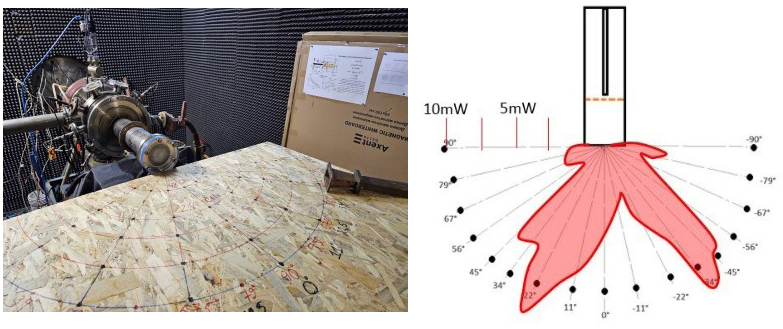}\\
  \caption{\emph{General view of the measuring stand for evaluating the directivity pattern and the type of it.}}\label{fig:image_label}
\end{figure}

According to the obtained intensity, the average power point was set on the amplitude scale ranging from $0^\circ$ to $180^\circ$. The maximum signal strength was recorded on the two petals of the pattern, located at the angles of $+22^\circ$ and $-34^\circ$. It was determined that this type of waveguide is characterized by a directivity pattern of the TM type with two main lobes.

\begin{center}
\fontsize{11}{14}\selectfont
\noindent \textbf{5. CONCLUSION}
\end{center}

Comprehensive measures were taken to reduce the impact of electromagnetic interference on diagnostic equipment and the electronic equipment of the power system of accelerator.
The use of cable shielding, grounding, separation of measuring equipment from the common ground, wireless measurement of power signals, waveguide-horn receiving antennas that cut off frequencies below 2.1 GHz, pyramidal absorbers and other measures have significantly reduced the level of electromagnetic interference. Conditions have been created for reliable and reproducible recording of signals in the gigahertz range.
The internal volume of the measuring anechoic chamber has been improved by separating the sources of useful signals from the sources of interference, represented by the controlled air dischargers and plasma guns, with a portable, collapsible, lightweight metal screen, which has reduced the intensity of electromagnetic interference by five times.
The spectra of the useful signal and the interference signal have been determined.
The method of determining the opening voltage using experimental current oscillograms was used, which allowed us to remove a capacitive voltage divider from the internal volume of the chamber, which in turn ensured the implementation of subsequent physical processes associated with the formation of useful signals.
A method was developed to experimentally determine the type of electromagnetic radiation pattern. The TM type of the pattern was determined on the DAEA DIN-2K.
A series of experiments were carried out that confirmed the stability of the main discharge circuit and plasma current switch, as well as the efficiency of the engineering solutions made to provide the interference protection. It was established that the current dynamics during the POS operation, as well as the amplitude values of the signals have a high level of repeatability. which is indicative of the stability of the unit.
The data obtained indicate the prospects of using the DAEA DIN-2K as a source of microwave radiation in scientific and applied research, and compliance with a set of measures taken to provide the interference protection ensures a reliable protection of the measuring equipment and a reduction in interference intensities.

\begin{center}
\fontsize{11}{14}\selectfont
\noindent \textbf{REFERENCES}
\end{center}

[1]	James Benford, John A. Swegle, Edl Schamiloglu. High Power Microwaves. CRC Press, 2015, 447 p.

[2]	Sohail Mumtaz, Han Sup Uhm, Eun Ha Choi. Progress in vircators towards high efficiency: Present state and future prospects. Physics Reports 1069 (2024), p.1-46.

[3]	D.V. Vinnikov, V.V. Katrechko, V.B. Yuferov, V.I. Tkachev. Plasma Guns of an Erosion Type with the Pulse-Periodic Gas-Metal Injection// Problems of Atomic Science and Technology. 2022, NO.6(142), p. 60-65. https://doi.org/10.46813/2022-142-060

[4]	D.V. Vinnikov, V.V. Katrechko, O.M. Ozerov, V.I. Tkachev, S.V. Marchenko  V.B. Yuferov, O.V. Manuilenko. Influence of the energy parameters of the primary circuit on the current characteristics of the DIN-2K accelerator. Problems of Atomic Science and Technology. "Series: Plasma Electronics and New Methods of Acceleration". 2023, NO.4, p. 36.

[5]	D.V. Vinnikov, V.V. Katrechko, O.V. Manuilenko, O.M. Ozerov, I.N. Onishchenko, V.I. Tkachev, V.B. Yuferov, S.V. Marchenko. Improving the functionality of the small-dimension accelerator DIN-2K with a plasma opening switch. Problems of Atomic Science and Technology. Series: "Nuclear Physics Investigations". 2024, NO.3(151), p. 60-66. https://doi.org/10.46813/2024-151-060

[6]	V.V. Katrechko, D.V. Vinnikov, O.V. Manuilenko, O.M. Ozerov, I.N. Onishchenko, V.I. Tkachev, V.B. Yuferov, S.V. Marchenko. Radiation in the microwave range at pulse accelerators with virtual cathode generator. Problems of Atomic Science and Technology. Series: "Nuclear Physics Investigations". 2024, NO.3(151), p. 79-83.

[7]	O.S. Druj, V.V. Yegorenkov, I.M. Onyshchenko, V.B. Yuferov. Plasma Dynamics in Accelerator with Plasma Opening Switch // Problems of Atomic Science and Technology. 2019, NO.6(124), p. 77-80. https://doi.org/10.46813/2019-124-077

[8]	O.S. Druj, V.V. Yegorenkov, A.V. Shchagin, V.B. Yuferov. Electron beam transport in dielectric tubes // East European Journal of Phisics. 2014, v.1, NO.1, p. 70-73.

[9]	Tsukasa Nakamura, et al. Output Evaluation of Microwave Pulse Emitted from Axially-Extracted Vircator with Resonance Cavity / Hasegawa, Jun (ed.). Tokyo Institute of Technology, Tokyo (Japan), Jan 2018, p. 55-60.

[10]	Sohail Mumtaz, Eun-Ha Choi. The Characteristics of the Second and Third Virtual Cathodes in an Axial Vircator for the Generation of High-Power Microwaves. Electronics 2022, NO.11(23), 973; https://doi.org/10.3390/electronics11233973

[11]	Se-Hoon Kim, Chang-Jin Lee, Wan-Il Kim, Kwang-Cheol Ko, "Investigation of an Axial Virtual Cathode Oscillator with an Open-Ended Coaxial Cathode", Journal of Electromagnetic Engineering and Science, vol.22, no.3, pp.265, 2022. DOI: 10.26866/jees.2022.3.r.86

[12]	Giacomo Migliore , Antonino Muratore , Alessandro Busacca , Pasquale Cusumano, Salvatore Stivala. Novel Configuration for a C-Band Axial Vircator With High Output Power. IEEE transactions on electron devices, VOL. 69, NO. 8, AUGUST 2022. P. -4579-4585. DOI: 10.1109/TED.2022.3184917

[13] Dr Frank Süli. Electronic Enclosures, Housings and Packages. 11 - Interference and shielding. 2019, Pages 499-526.

[14]	Experimental study of plasma opening switch in electron accelerator with inductive energy storage. D.V. Vinnikov, O.M. Ozerov, V.V. Katrechko, V.I. Tkachov, S.V. Marchenko, V.B. Yuferov, O.V. Manuilenko , I.N. Onishchenko. Problems of Atomic Science and Technology. 2025. NO.4(158), P.3-10. DOI: 10.46813/2025-158-003

[15]	Conditions ensuring maximum voltage multiplication factors for the din-2k accelerator. D.V. Vinnikov, O.M. Ozerov, V.V. Katrechko, V.I. Tkachov, S.V. Marchenko, B.O. Brovkin, V.B. Yuferov, O.V. Manuilenko, I.N. Onishchenko. Problems of Atomic Science and Technology. 2025. NO.1(155), p 69-74. DOI: 10.46813/2025-155-069.

[16]	Ware K. D., Filios P. G., Gullickson R. L., Rowley J. E., Schneider R. F., Summa W. J., Vitkovitsky I. M. 1997 Inductive energy technology for pulsed intence X–ray sources. IEEE Trans. Plasma Sci. NO.25(4), 160–168. CrossRefGoogle Scholar

[17]	Yinfeng Zhou, Ming Chen, Yu Tian, Xun Zhou, Pei Shen. Study on EMC and OTA Anechoic Chamber Compatibility. Journal of Electronic Research and Application, 2025, Volume 9, Issue 4., p.87-96. DOI: 10.26689/jera.v9i4.11450

[18]	O.V. Manuilenko, I.N. Onishchenko, A.V. Pashchenko, I.A. Pashchenko, V.B. Yuferov. Current Flow Dynamics in Plasma Opening Switch // Problems of Atomic Science and Technology. 2021, NO.4(134), p. 6-10. https://doi.org/10.46813/2021-134-006

[19]	O.V. Manuilenko, I.N. Onishchenko, A.V. Pashchenko, I.A. Pashchenko, V.A. Soshenko, V.G. Svichensky, V.B. Yuferov, B.V. Zajtsev. Magnetic Field Dynamics in Plasma Opening Switch // Problems of Atomic Science and Technology. 2021, NO.6(136). p. 61-66. https://doi.org/10.46813/2021-136-061

[20]	O.V. Manuilenko, I.N. Onishchenko, A.V. Pashchenko, I.A. Pashchenko, V.B. Yuferov. Plasma and Magnetic Field Dynamics in POS: PIC simulations // Problems of Atomic Science and Technology. 2022, NO.6(142). p. 55-59. https://doi.org/10.46813/2022-142-055

[21]	Schumer, J. W., Swanecamp, S. B., Ottinger, P. F., Commisso, R. J., Weber, B. V., Smithe, D. N. Ludeking, L. D. MHD-to-PIC transition for modeling of conduction and opening in a plasma opening switch. IEEE Trans. 2001 Plasma Sci. 29 (3), 479–493.CrossRefGoogle Scholar

[22]	Ernesto Neira, Felix Vega, Chaouki Kasmi, Fahad AlYafei, "Power Capabilities of Vircators: A Comparison between Simulations, Experiments, and Theory", 2020 IEEE 21st International Conference on Vacuum Electronics (IVEC), pp.313-314. 2020. DOI: 10.1109/IVEC45766.2020.9520518

[23]	Sunil Kanchi1, Rohit Shukla, Archana Sharma. Simulation and experimental results of plasma opening switch operation with inductive and electron beam diode loads in different plasma regimes. Physica Scripta, Volume 99, Number 2. 2024. 025614. DOI 10.1088/1402-4896/ad1d9d.

\end{multicols}

\end{document}